\documentclass[aps,prb,twocolumn,groupedaddress]{revtex4} 
\usepackage{amssymb}
\usepackage{graphicx}
\usepackage{epsfig}
\usepackage{float}
\usepackage{amsmath}

\begin{document}

\title{Fermi Surface Nesting and the Origin of the Charge Density Wave in
NbSe$_2$}
\author{M.D. Johannes, I.I. Mazin, C.A. Howells}
\affiliation{Code 6391, Naval Research Laboratory, Washington, D.C. 20375}

\begin{abstract}
We use highly accurate density functional calculations to study the band
structure and Fermi surfaces of NbSe$_2$. We calculate the real part of the
non-interacting susceptibility, $\Re\chi_0(\mathbf{q})$, which is the
relevant quantity for a charge density wave (CDW) instability and the
imaginary part,$\Im\chi_0(\mathbf{q})$, which directly shows Fermi surface
(FS) nesting. We show that there are very weak peaks in $\Re\chi_0(\mathbf{q}
)$ near the CDW wave vector, but that no such peaks are visible in $
\Im\chi_0(\mathbf{q})$, definitively eliminating FS nesting as a factor in
CDW formation. Because the peak in $\Re\chi_0(\mathbf{q})$ is broad and
shallow, it is unlikely to be the direct cause of the CDW instability. We
briefly address the possibility that electron-electron interactions (local
field effects) produce additional structure in the total (renormalized)
susceptibility, and we discuss the role of electron-ion matrix elements.
\end{abstract}

\maketitle

In 1964 V.L. Ginzburg proposed excitonic superconductivity in quasi-2D
structures\cite{VL} composed of metal layers sandwiched between insulating layers. By
that time, only two layered materials were known to be
superconducting,\cite{PDT,EREPL63} PdTe$_{2}$ and NbSe$_{2}$.  In the four decades
since then, NbSe$_{2}$ and isostructural selenides have been intensively investigated
(close to 1.5 thousand papers published to date). However, the main interest in this
compound has shifted from the fact that it is a layered material (hundreds of layered
superconductors are now known), to the existence of a nearly-commensurate charge
density wave (CDW)  \cite{SDW-1} instability and it's possible interplay with the
superconductivity that sets in at a lower temperature. We mainly forego discussion of
the interesting issues surrounding the superconducting state, its origin, and its
relationship to the CDW state, and instead concentrate on the mechanism behind the CDW
transition itself.

The first electronic structure calculations for NbSe$_2$ were presented by Mattheiss in
1973 \cite{LFM73}. Using a non-self-consistent potential he was able to produce a band
structure with basic features in reasonable agreement with more recent self-consistent
calculations \cite{RCPM+94,TSTF+99,KROS+01}, but which showed only two bands crossing
the Fermi energy (it is now known that there are three), and underestimated the energy
depth of a saddle point at $\sim \frac{1}{2}$ $\Gamma K$. Fermi surfaces based on this
band structure led to early suggestions that the CDW transition was driven by nesting
\cite{SDW-1}, an assumption that has carried through to the present time. The nearness
of the saddle point to the Fermi energy (E$_F$) led Rice and Scott \cite{TMR75} to
argue that CDW formation was driven, not by Fermi surface (FS) nesting in the
conventional sense, but rather by saddle points lying within $k_BT_{CDW}$ of E$_F$ and
separated by the CDW wavevector, $ \mathbf{Q}_{CDW} =(\frac{1}{3},0,0)$. A significant
amount of effort has been spent on resolving the `controversy' between the nesting and
saddle point theories for NbSe$_2$ and for related CDW compounds such as 2H-TaSe$_2$,
1T-TaSe$_2$, TaS$_2$ and others, but no specific feature that would give rise to an
instability at $\mathbf{{Q}_{CDW}}$ has been convincingly isolated. As early as 1978,
Doran \textit{et al} \cite{NJD+78,NJD78} used a simple one-band tight-binding model to
show that the susceptibility has no sharp peaks of electronic origin and made a
rudimentary calculation of the electron-ion matrix elements that suggested strong
coupling at $\mathbf{Q}_{CDW}$. Nonetheless, in the more modern era of
self-consistent calculations, the FS nesting and saddle point hypotheses continue to be
debated and FS nesting is cited as a contributing factor in CDW formation in almost
every paper that addresses the issue. In this paper we report a quantitative analysis
of the non-interacting susceptibility, $\chi_{0}(\mathbf{q},\omega=0)$, based on first
principles density functional theory (DFT) calculations, and show that not only is the
structure in momentum space extremely weak, but that FS nesting does not play any role
at all and can be definitively ruled out as a cause for CDW formation. Instead, we
argue, the instability must be due to the electron-ion interaction, or possibly from
local field effects. The structure of the paper is as follows: In Section I, we lay out
the basic mathematical framework for dielectric calculations, and discuss the various
approximations we employ. In Section II, we present and discuss the calculated band
structure and Fermi surfaces and show that there is good agreement with experiment. In
Section III, we show the real and imaginary parts of the non-interacting
susceptibility, and discuss the origins of prominent features, emphasizing the lack of
a strong peak at the CDW wave vector and the irrelevance of FS nesting.

\section{General Theory}

A material is unstable toward the formation of a CDW at wave vector $\mathbf{
Q}$ when the response function (inverse electronic dielectric function)
fails to satisfy the stability condition \cite{Dolgov1},

\begin{equation}
\epsilon^{-1}(\mathbf{Q},0) =1+V_{C}(Q)\chi(\mathbf{Q},0) \leq 1
\label{fchi}
\end{equation}
where $V_{C}(q)=4\pi e^{2}/q^{2}$ is the Coulomb potential, and the
susceptibility $\chi(\mathbf{q},0)$ is normally negative. A dielectric constant
within the range $0 \leq \epsilon(\mathbf{Q},\omega) < 1 $ therefore signals
a CDW instability and, in conjunction with Eq. \ref{fchi}, indicates that
the latter range is accessible only from the lower boundary, \textit{i.e}
when $\chi(\mathbf{Q},\omega)$ diverges. In the simplest approximation, $
\chi(\mathbf{q},\omega)$ is calculated by ignoring the exchange part of the
electron-electron interaction in favor of Hartree-like terms only. This is
known as the random phase approximation (RPA) \cite{PNDP58} and relates $
\chi $ to the non-interacting susceptiblity, $\chi_0$: 
\begin{equation}
\chi_{RPA}(\mathbf{q},\omega)=\chi_{0}(\mathbf{q},\omega)/[1-V_{C}(q)
\chi_{0} (\mathbf{q},\omega)],  \label{RPA}
\end{equation}
where $\chi_0$ is 
\begin{equation}
\chi_{0}(\mathbf{q},\omega)=\sum_{ll^{\prime },\mathbf{k}}\frac{
f(\varepsilon _{l\mathbf{k}})-f(\varepsilon_{l^{\prime }\mathbf{k+q}})}{
\varepsilon_{l\mathbf{k} }-\varepsilon_{l^{\prime }\mathbf{k+q}
}-\omega-i\delta}\left |\left\langle l\mathbf{k} |e^{i\mathbf{qr}}|l^{\prime
},\mathbf{k+q}\right\rangle \right |^{2}.  \label{Re}
\end{equation}
Here $f(\varepsilon)$ is the Fermi function and the matrix elements $
\left\langle l\mathbf{k}|e^{i\mathbf{qr}}|l^{\prime },\mathbf{k+q}
\right\rangle $ are between Bloch functions, $|l\mathbf{k}\rangle = u_{l
\mathbf{k}}(\mathbf{r})e^{i\mathbf{kr}}$. If these do not have much
structure as a function of \textbf{q}, one can set them uniformly to unity,
yielding the following expression for the imaginary part of $\chi_0$:

\begin{equation}
\lim_{\omega \rightarrow 0}\Im \chi _{0}(\mathbf{q},\omega )/\omega =\frac{
\Omega }{(2\pi )^{3}}\sum_{ll^{\prime }}\int \frac{dL_{\mathbf{k}}}{|\mathbf{
v}_{l\mathbf{k}}\times \mathbf{v}_{l^{\prime },\mathbf{k+q}}|},
\label{nesting}
\end{equation}

where the line $L_{\mathbf{k}}$ is the intersection of the two surfaces
defined by $\varepsilon _{l\mathbf{k}}=0$ and $\varepsilon _{l^{\prime }
\mathbf{k+q}}=0$ (the Fermi energy is set to zero). Eq. \ref{nesting}
correctly suggests that a Fermi surface nesting, that is, a Fermi surface
topology such that the length of the line $L_{\mathbf{k}}$ is particularly
large at some \textbf{Q}, should lead to a peak in $\lim_{\omega \rightarrow
0}\Im \chi _{0}(\mathbf{Q},\omega )/\omega .$ This is correct, but misses
several important points. First, as Eq. \ref{nesting} indicates, a strong
maximum also requires small and/or nearly parallel $\mathbf{v}_{l\mathbf{k}}$
and $\mathbf{v}_{l^{\prime },\mathbf{k+Q}}$ along $L_{\mathbf{k}}.$ Second,
it is the real, and not the imaginary part of the static susceptibility that
defines a CDW instability (at $\omega =0,\Im \chi _{0}$ vanishes). In
contrast to $\Im \chi _{0}$, which indeed depends only on electronic
characteristics near the Fermi surface, $\Re \chi _{0}$ collects information
from an energy window of the order of the band width (which follows from
Eq. \ref{Re}, taking into account that $\int dE/E$ diverges at large
arguments).

Finally, one should keep in mind that a divergence of the electronic susceptibility
signals an instability of the system with respect to spontaneous formation of a CDW
\textit{even if the nuclei are clamped to their high-symmetry positions}. Of course, if
released, the nuclei would shift as well. However, divergence of $\chi$ is a
substantially more severe criterion than the soft mode condition, which is that the
frequency of a particular phonon mode at some wave vector becomes zero. This latter
condition defines an instability toward the freezing in of a particular periodic
pattern of ionic displacement, which is the only CDW that is actually observed in real
systems.

Phonon frequencies are related to the dielectric function via the
so-called Pick-Cohen-Martin formula \cite{pick}. Here we give this formula in a
simplified form for an elemental solid:

\begin{eqnarray} D_{\alpha \beta }(\mathbf{q)} &\mathbf{=}&\tilde{D}_{\alpha \beta }(\mathbf{ q)-}\tilde{D}_{\alpha \beta
}(0\mathbf{),} \notag \\ 
\tilde{D}_{\alpha \beta }(\mathbf{q)} &\mathbf{=}&\sum_{\mathbf{G},\mathbf{G} ^{\prime
}}|\mathbf{q+G}|^{2}V_{C}(\mathbf{q+G)}(\mathbf{q+G)}_{\alpha }\times \\ 
& &\epsilon
^{-1}(\mathbf{q+G},\mathbf{q+G}^{\prime })\mathbf{(q+G}^{\prime })_{\beta }V_{C}(\mathbf{q+G}^{\prime }\mathbf{),} \notag \\  
& & \notag
\label{PCM}
\end{eqnarray} 

where $D_{\alpha \beta }(\mathbf{q)}$ is the dynamical matrix and $\alpha $, $\beta $
are Cartesian indices. It is clear that an eigenvalue of this matrix can soften and
eventually become zero even if the macroscopic dielectric function\cite{SLA62,NW63}:  
\begin{equation} \epsilon (\mathbf{q})=1/[\epsilon^{-1}
(\mathbf{q+G},\mathbf{q+G}^{\prime })]_{00} \label{AW} \end{equation} does not diverge.  
Here and above, the dielectric function $\epsilon
^{-1}(\mathbf{q+G},\mathbf{q+G}^{\prime })$ differs from $\epsilon ^{-1}_{RPA}$ in that
it includes the so-called local field effects: Umklapp processes and the
exchange-correlation interaction between electrons. Formally, it is written (neglecting for simplicity the 
band indices) as:

\begin{widetext} \begin{eqnarray} \epsilon ^{-1}(\mathbf{q+G},\mathbf{q+G}^{\prime
})=&\delta _{\mathbf{GG}^{\prime }}&+\sum_{\mathbf{G}_{1}}V_{C}(\mathbf{q+G)}\chi
_{0}(\mathbf{q+G,q+G }_{1})\otimes \notag\\ & [\delta
_{\mathbf{G}_{1}\mathbf{G}^{\prime}}& -\sum_{\mathbf{G} _{2}}\{\delta
_{\mathbf{G}_{1}\mathbf{G}_{2}}V_{C}(\mathbf{q+G}_{1})-I_{XC}(
\mathbf{q+G}_{1},\mathbf{q+G}_{2})\}\chi _{0}(\mathbf{q+G}_{2}\mathbf{,q+G}^{\prime })]
^{-1} \end{eqnarray} \end{widetext} (in these expressions, matrix inversion is with
respect to the reciprocal lattice vector indices), where $I_{XC}(\mathbf{r,r}^{\prime
})=\delta ^{2}E_{XC}/\delta \rho (\mathbf{r})\delta \rho (\mathbf{r^{\prime } })$is the
exchange correlation kernel [note that in the LDA $I_{XC}(\mathbf{ r-r^{\prime
}})=\delta (\mathbf{r-r^{\prime }})I_{XC}(\mathbf{r})$, and thus
$I_{XC}(\mathbf{q+G,q+G^{\prime }})=I_{XC}(\mathbf{G-G^{\prime }})]$.

Therefore, several separate questions must be posed: 1) Does the geometry of the Fermi surface provide nesting in the sense of a
strong maximum of Eq.\ref{nesting} ? 2) Is there a peak in the real part of $\chi _{0}$, as defined in Eq.\ref{Re} at the wave
vector corresponding to the observed CDW and if so, is it related to any nesting-derived structure in $\Im \chi(\mathbf{q},0) ?$ 
3) Is there a
divergence in $\epsilon (\mathbf{q})$ as defined by Eq.\ref{AW} ? 4) If the previous answer is no, is there a 
soft mode (zero
eigenvalue) in Eq. 5?

Below we present numerical density functional calculations that provide answers to the first two questions and supply
information allowing for reasonable conjecture about answers to the last two. The substance of our findings is that there
is only weak nesting in this system, and at the \textquotedblleft wrong\textquotedblright\ wave vector. While $\Re \chi
_{0}$ does have a broad maximum at the CDW wave vector, it is likely too small to account for the CDW instability.
Lacking full linear response calculations, we cannot exclude with certainty an electronic instability in $\epsilon
(\mathbf{q})$ (Eq. \ref{AW}), due to local field effects (exchange-correlation and Umklapp) beyond plain RPA, but it is
more likely that the instability appears only after electron-ion interactions are explicitly taken into account (Eq. 5).  
Experiment further supports our contention as only the phonon corresponding to the actual CDW is observed to soften, 
whereas, if $\epsilon (\mathbf{q})$ were to diverge at some $\mathbf{q}$, {\it all} phonons would soften at this vector.

To illustrate the last point we note that $\chi _{RPA}$ itself can have a peak, but obviously cannot diverge
(Eq. \ref{RPA}).  Furthermore, including the exchange-correlation interaction between electrons within DFT,
but without Umklapp processes, gives \begin{equation} \chi _{DFT}(\mathbf{q},0)= \frac{\chi
_{0}(\mathbf{q},0)}{[1+\{V_{C}(q)-I_{XC}(q)\} \chi _{0}(\mathbf{q},0)]} \label{ex} \end{equation} This
expression cannot diverge either. Finally, it can be shown that, generally speaking, the Umklapp processes
tend to \textit{lower} the electronic dielectric constant\cite{Dolgov1}. Since $\epsilon (\mathbf{q} ,\omega
)$ can become unstable only by passing from a negative quantity to a positive one, there can be no CDW
formation at any value of $\mathbf{q}$. Non-local correlation effects neglected by DFT may be instrumental in
creating an instability beyond Eq. \ref{ex}, but it is physically more likely that the electron-ion matrix
elements, which can naturally have strong $\mathbf{q}$-dependence, account for the CDW instability through 
their effect on the dynamical matrix. In fact, this is consistent with some early conjectures based on
simplified tight-binding calculations \cite{NJD78,KMEA83}. In these works, a strong phonon softening was
observed at the correct wave vector, but only when electron-ion interactions were accounted for. The authors
of both works conclude, correctly, that this softening is not simply related to FS geometry, but stay short
of excluding weak FS nesting as an essential factor. As we will show, no relevant nesting exists in
NbSe$_{2}$, and, therefore, it cannot play any role in the CDW instability.

\section{Electronic Structure}

\begin{figure}[tbp]
\includegraphics[width =0.85\linewidth]{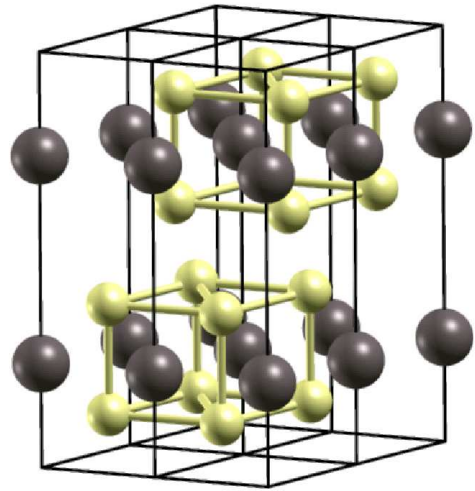}
\caption{The crystal structure of NbSe$_2$, showing the two distinct layers
that make up the unit cell. Each Nb ion sits in a hexagonal mirror plane
with three Se neighbors above and below, forming a network of distorted
edge-sharing octahedra.}
\label{struct}
\end{figure}

Our calculations were carried out using a full-potential, augmented plane wave plus local orbital (APW+lo) scheme
\cite{Wien2k}. The exchange-correlation potential was approximated by a local density approximation (LDA) parameterization
\cite{JPP92} and the muffin-tin radii were set to 2.3 (Nb) and 2.2 (Se)  with a value of 7.0 for RKmax. We found small but
important differences between calculations with and without the spin-orbit interaction. Except where explicitly noted, we
include spin-orbit coupling using a second variational method for all calculations. The structure shown in Fig. \ref{struct}
displays the unit cell of NbSe$_{2}$. It is comprised of two formula units arranged in layers, each layer containing a
hexagonal Nb plane sandwiched between two shifted hexagonal Se planes. We used the experimentally measured lattice constants
\cite{MMPDD+72}, (a=3.44 $\mathring{A}$ and c=12.55 $\mathring{A}$), which agrees well with other reports
\cite{str1,str2,str3} and we relaxed the Se ions to their lowest energy positions along the $c$ axis ($z$=0.1183$c$). This
position is in good agreement with all experimental reports ($z=0.116-0.118c)$\cite{MMPDD+72,str1,str2}, except Ref.  
\onlinecite{str3}, which gives $z=0.125c.$ Our relaxation resulted in a Se-Se interplanar distance of 3.57 $\mathring{A}$
which can be compared with the Nb-Se intralayer distance of 2.59 $\mathring{A}$. Based on these separations, we expect an
anisotropic band structure with rather strong interplanar coupling. To calculate the non-interacting susceptibility of Eq.
\ref{Re}, we used a direct summation over a very fine mesh of 73$\times $73$\times $7 $ \sim $ 40,000 $k$ points in the BZ,
and a Fermi temperature smearing of T= 10 meV.

\subsection{Band Structure}

\begin{figure}[tbp] \includegraphics[width = 0.95\linewidth]{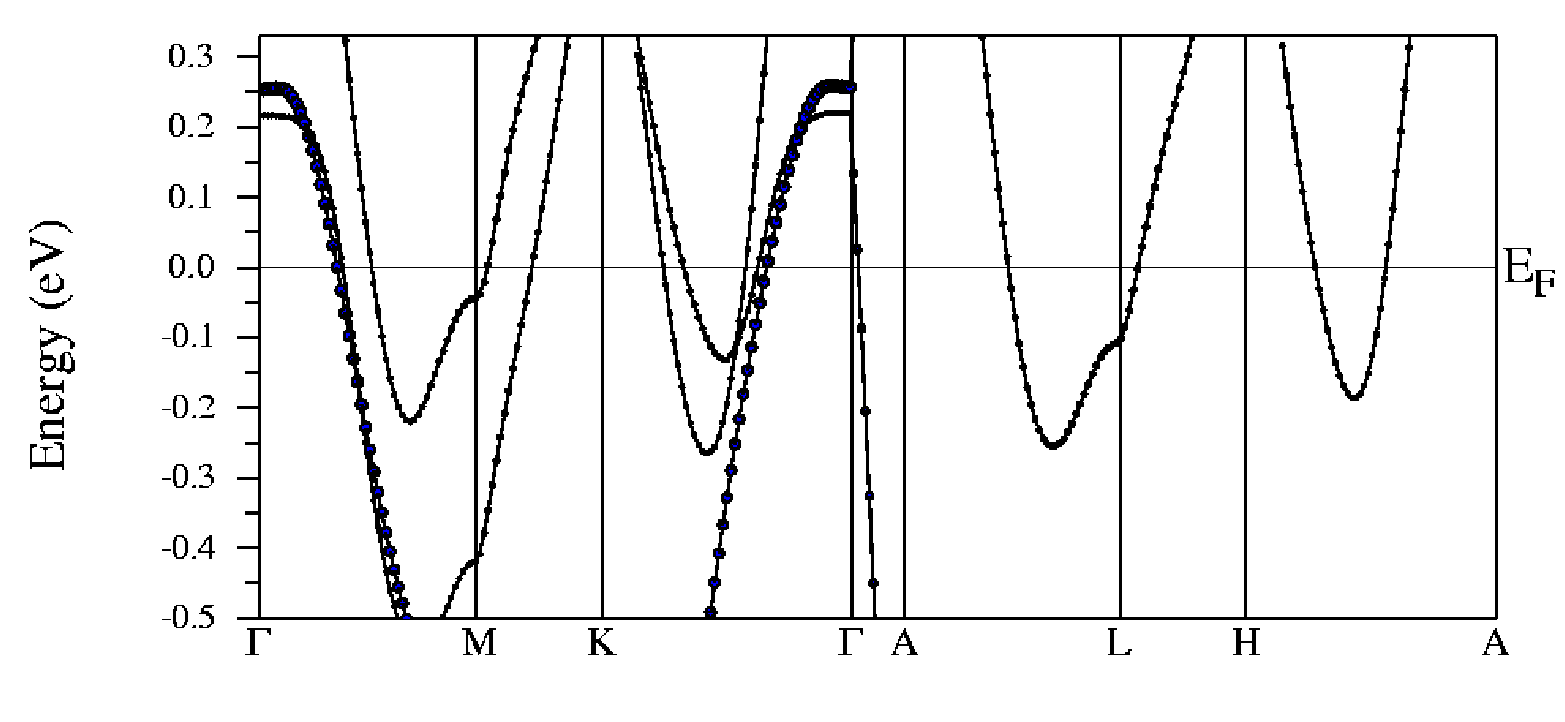} \includegraphics[width =
0.95\linewidth]{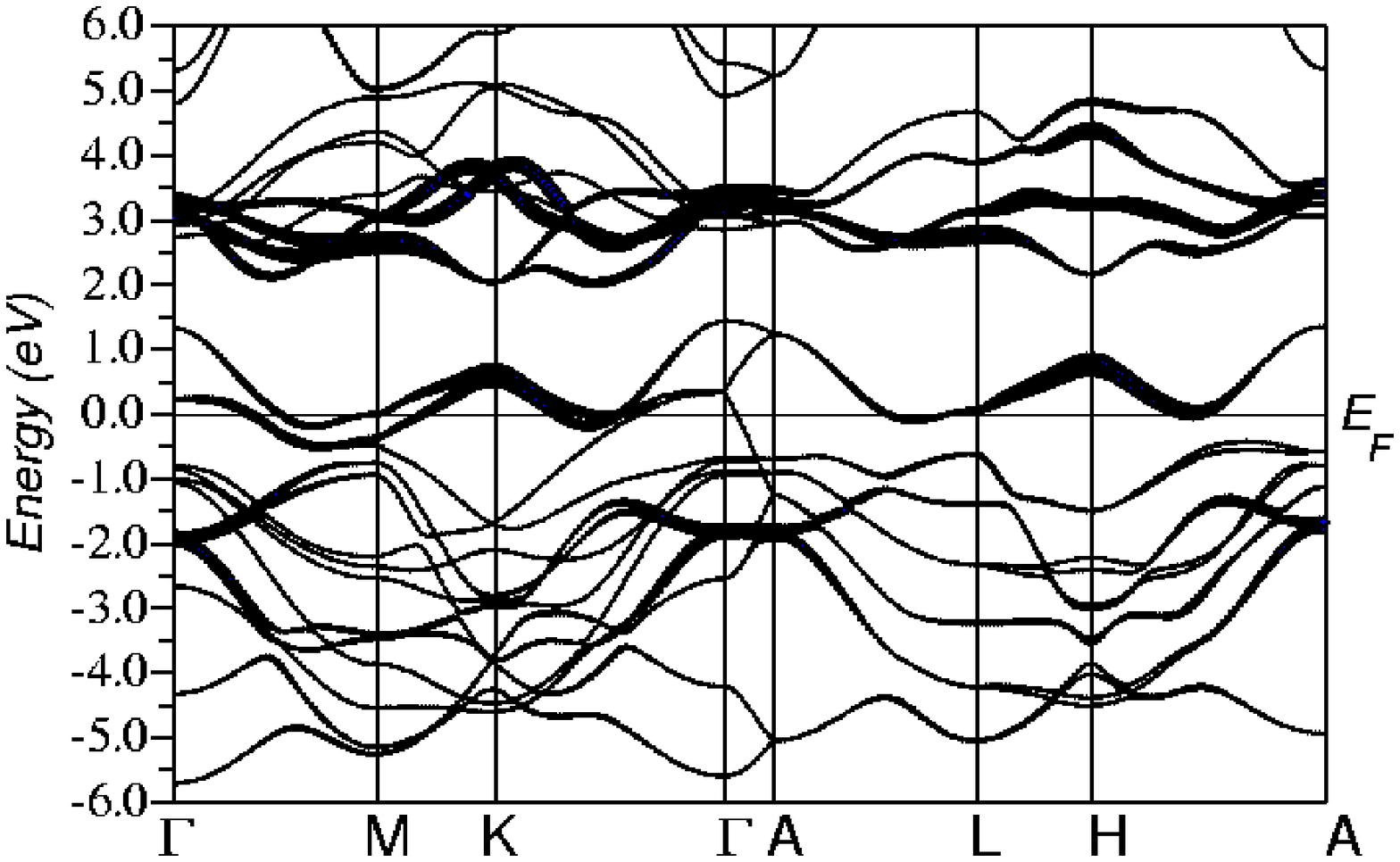} \caption{(color online) The band structure of NbSe$_2$. {\it Top:} A closeup of the bands around
the Fermi energy, neglecting spin-orbit coupling.  The highlighted bands are of Se $p_z$ character and cross
the Nb bands without hybridization.  {\it Bottom:} The full band structure, including the spin-orbit
interaction which relieves the Nb band degeneracy in the $k_z = \pi/c$ plane.  The highlighted bands are of
combined $x^2-y^2/xy$ character} \label{bs} \end{figure}

Our band structure (Fig. \ref{bs})  exhibits some differences from previous full potential calculations,
\cite{RCPM+94,KROS+01} giving rise to differences in the resulting Fermi surfaces. The discrepancies are most likely due to
a large difference in Se positions: $z$=0.134$c$ in Refs.  \onlinecite{RCPM+94,KROS+01} compared to our relaxed height of
$z$=0.118c. We were not able to identify an experimental paper reporting such high Se positions.

\begin{figure} \includegraphics[width = 0.95\linewidth]{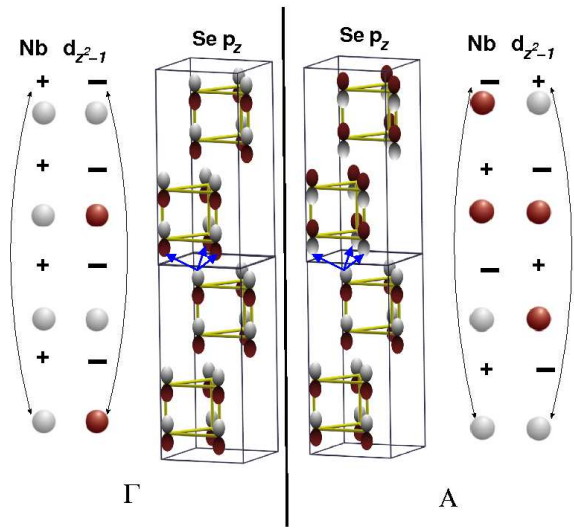} \caption{(color online) A schematic showing
the origin of the splitting of the antibonding $p_z$ band and the Nb $d_{z^2-1}$ band at two different $k_z$
positions. The Nb ions are omitted from the structural graphic for clarity and the $d_{z^2-1}$ orbitals have
been idealized as spheres. {\it Left side:} At the $\Gamma$-pt, Nb $d_{z^2-1}$ orbitals are either fully
bonding or fully antibonding, and the antibonding $p_z$ orbitals are antibonding even between cells (blue
arrows).  {\it Right side:} At the $A$-pt, there is no overall energy difference between bonding and
antibonding Nb $d_{z^2-1}$ configurations and the antibonding $p_z$ orbitals are now bonding between cells.  
Plus signs indicate bonding and minus signs antibonding - the long black arrows connect the topmost orbital
to the one that would sit above it if the cells were repeated endlessly.} \label{GA} \end{figure}

We observe two classes of bands near the Fermi energy - the first is the fully antibonding Se $p_{z}$ band,
which, by parity, does not hybridize with the $z-$even d-orbitals of Nb ($d_{x^{2}-y^{2}},$ $ d_{xy},$ and
$d_{z^{2}-1}$) in the $k_{z}$=0 plane, and does not hybridize with the $z-$odd Nb $d$ orbitals in the
$k_{z}=\pi /c$ plane.  In Fig. \ref{bs}a, a close-up of the band structure without spin-orbit interactions is
shown.  The $p_z$ band is highlighted with filled circles and can be seen crossing and re-crossing the Nb band in the
$\Gamma-K-M$ plane without mixing, despite a nearly perfect energetic degeneracy.  In the $A-L-H$ plane, the
range of the figure no longer includes the $p_z$ band which has dropped nearly 2 eV in energy. This rather dramatic
energy lowering can be traced to bonding-antibonding interactions between unit cells.  The width of the Se
$p$ band complex (-6.0 eV to 0.5 eV)  is determined by the energy difference between the bonding and
anti-bonding configurations of the four $p_z$ orbitals along the $c$-axis within the (two layer) unit cell
and between the cells themselves.  In the $k_z$=0 plane, antibonding $p_z$ orbitals in adjacent unit cells
are antibonding with each other, pushing the energy of the single fully anti-bonding $p_z$ band up to the
Fermi level.  In the $k_{z}=\pi /c$ plane, the relationship between adjacent-cell $p_z$ orbitals becomes
bonding, the band is lowered, and the Fermi crossing is removed. Fig.  \ref{GA} shows the antibonding
configuration of the $p_z$ orbitals at the $\Gamma$ and $A$ points, with arrows indicating the crucial
interactions between cells.

The second class of bands consists of one Nb $d$ band, which, like the $p_z$ orbitals, is split in the
$k_z$=0 plane, by bonding-antibonding interactions both within and between unit cells.  In Fig. \ref{GA},
spheres representing the Nb $d_{z^2-1}$ orbital (or equally the other $z$-even orbitals, $x^2-y^2$ and
$xy$) are shown in fully bonding and fully antibonding configurations at the $\Gamma$ point.  By
contrast, in the $k_{z}=\pi /c$ plane ($A$ point shown), the bonding-antibonding energy difference
arising from orbital orientations within the unit cell is counteracted by interactions between cells and
the band therefore remains degenerate (see Fig. \ref{bs}a).  In Fig. \ref{bs}b, the full band structure,
including spin-orbit, is shown.  The spin-orbit coupling allows for $p_z$-Nb mixing along $\Gamma-K-M$
and removes the strict degeneracy of the Nb bands at $k_{z}=\pi /c$. It is tempting to identify the
latter bands as simply Nb $d_{z^{2}-1}$ orbitals,
but is misleading: symmetry arguments, confirmed by the numerical decomposition of the LAPW
bands, predict that this band should have atomic $d_{z^{2}-1}$ character along the $\Gamma $A line, but
$d_{x^{2}-y^{2}}+d_{xy}$ character along the KH line and mixed character elsewhere. This can be seen
clearly in Fig.\ref{bs}b the where the $d_{x^{2}-y^{2}}+d_{xy}$ bands are highlighted and are clearly
dominant near the Fermi level along some symmetry directions. On can, of course, enforce mapping of this
band structure onto a one-band $dz2-1$ TB model \cite{RLB+05},
but this does not reflect the character decomposition of the LAPW bands
nor does it mean that this band is physically derived from actual Nb
$d_{z^2-1}$ orbitals.

\subsection{Fermi Surfaces}

\begin{figure}[tbp]
\includegraphics[width=0.95 \linewidth]{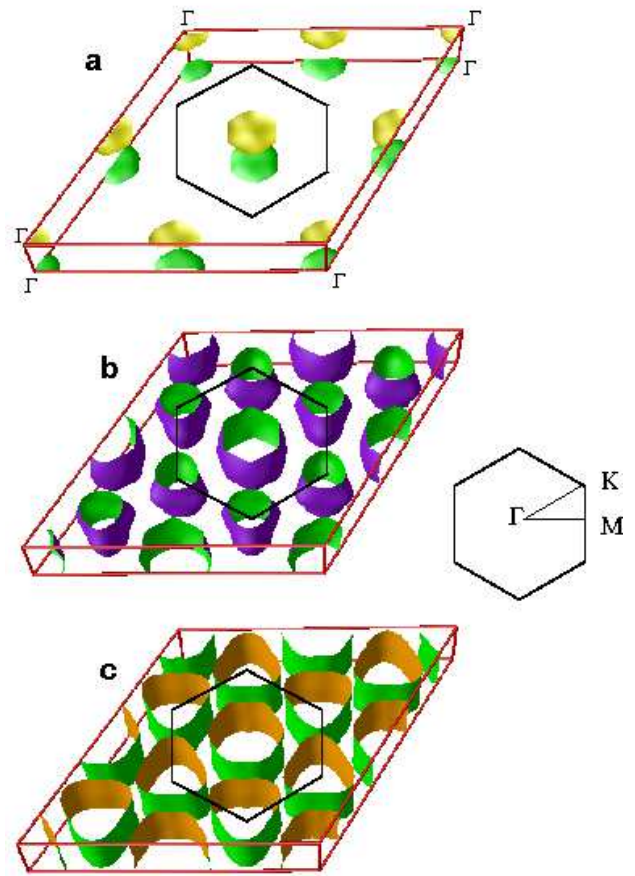}
\caption{The three Fermi surfaces of NbSe$_2$ in an extended BZ scheme with
the conventional BZ indicated as a hexagon in the center of each figure. 
\textbf{a}) The small Se-derived (band 1) pancake surface around the $\Gamma$
-point. \textbf{b}) The bonding (band 2) Nb-derived Fermi surface with
strong warping along $k_z$. \textbf{c}) The nearly 2D anti-bonding (band 3)
Nb-derived Fermi surface. The $\Gamma-K-M$ plane is at the top and bottom of
each figure and the $A-L-H$ plane cuts through center.}
\label{FS}
\end{figure}

In Fig. \ref{FS} our three calculated Fermi surfaces are displayed in an
expanded (four total zones) Brillouin zone scheme. These surfaces are
significantly different from early calculations, \cite{LFM73,CYF74,GWAMW76}
most notably in the existence of a third band crossing $E_{F}$. On the other
hand, they are quite similar to the sketches of Ref. \onlinecite{RCPM+94},
with the difference that we designate all three surfaces as
\textquotedblleft hole-like\textquotedblright\ because the filled states are
external to the drawn surfaces, whereas in Ref. \onlinecite{RCPM+94} the
third surface is called \textquotedblleft electron-like\textquotedblright ,
presumably because the band is more than half-filled. Below we compare each
FS to experimental data from angle-resolved photoemission spectroscopy
(ARPES) and de Haas van Alphen (dHvA) quantum oscillation measurements.

Interlayer coupling breaks the degeneracy of the highest unfilled Nb $d$ band and gives rise to two concentric cylindrical
Fermi surfaces around the $ \Gamma-A$ and $K-L$ lines of the BZ. The surfaces derived from the higher (anti-bonding) energy
band (Fig. \ref{FS}c) are almost dispersionless along the $k_z$ direction, while those derived from the lower (bonding)
energy band (Fig. \ref{FS}b) are strongly warped. Neither of these have been seen with dHvA, which requires very pure
samples to detect large orbits, but they have been consistently seen by ARPES measurements
\cite{TSTF+99,KROS+01,TYTK+01,TKTY+02,TVAVF+04} which are in very good general agreement with calculation. We will use the
measurements of Rossnagel \textit{et al} \cite{KROS+01}, who investigated directions both parallel and perpendicular to the
a-b plane, for specific comparison with our surfaces. The observed Fermi surfaces derived from the higher Nb band (Fig.
\ref{FS}b) show less $k_z$ dispersion than in our calculations and consequently the pockets around $K-L$ remain rounded
along their full length rather than becoming sharply triangular as ours do. The outer pockets of the lower Nb band (Fig.
\ref{FS}c) are not warped, consistent with our results, but the $ K-L$ centered triangular pockets again appear to be more
rounded. Overall there is substantial agreement between ARPES measurements and theory, particularly in terms of the observed
placement and $k_F$ of the Nb-derived FS pockets - the essential elements of interest for Fermi surface nesting.

\begin{table}[tbp]
\begin{tabular}{|l|c|}
\hline
& dHvA frequency (T) \\ \hline
\textbf{Band 1} &  \\ 
(001) & 4498.4 \\ 
(1$\bar{2}$0) & 472.1 \\ 
\textbf{Band 2} &  \\ 
$\Gamma-A$ ($\Gamma$) & 6269.9 \\ 
$\Gamma-A$ ($A$) & 8605.6 \\ 
$K-L$ ($\Gamma$) & 3199.0 \\ 
$K-L$ ($A$) & 5110.9 \\ 
\textbf{Band 3} &  \\ 
$\Gamma-A$ ($\Gamma$) & 9362.6 \\ 
$\Gamma-A$ ($A$) & 9046.6 \\ 
$K-L$ ($\Gamma$) & 7807.2 \\ 
$K-L$ ($A$) & 7085.2 \\ \hline
\end{tabular}
\caption{The calculated dHvA frequencies of NbSe$_2$. Bands two and three
have two separate surfaces, designated in the Table by the axis that runs
through their centers $\Gamma-A$ or $K-L$. The two extremal orbits for each
surface in a (001) field occur in the $\Gamma-K-M$ and $A-L-H$ planes,
indicated by $\Gamma$ and $A$ respectively in the Table. For
the Se pancake surface, the frequency was also calculated for an in-plane
field direction (1$\bar{2}$0).}
\label{dhva}
\end{table}

The small pancake surface in Fig. \ref{FS}a is derived from a Se $4p$ band, and has an extremely small width
along the $k_z$ direction. Very early dHvA measurements \cite{JEG76} saw this surface, but misattributed it
by comparing to a band structure \cite{LFM73} that didn't show the Se band crossing the Fermi energy. Later
measurements by Onuki \textit{et al} \cite {YOIU+91} confirm the existence of these surfaces, though the
authors were still apparently unaware of the calculated Se-derived Fermi surface and therefore report a
discrepancy with theory. The most recent dHvA experiments \cite{RCPM+94} clearly detect and correctly
identify the smallest surface, obtaining good quantitative agreement with the earlier works.  ARPES
measurements see a weak signal of the smallest surface \cite{KROS+01,TYTK+01} with an in-plane area that
agrees well with dHvA measurements, but with greatly exaggerated perpendicular dimension.  Our calculated Se
Fermi surface is larger both in- and out-of-plane with respect to dHvA measurements, but shows a smaller
perpendicular width than ARPES, which is not expected to be as accurate in this direction.  For comparison
with available experimental data, our calculated the dHvA frequencies for the Se-derived pocket as well as
all the other Fermi surfaces is shown in Table \ref{dhva}.

Note that our calculated cross-sectional areas are noticeably different from previous work by Corcoran {\it
et al}. This can be partially ascribed to the different Se position in their calculations. However, there
were also probably too few $k$-points to ensure an accurate area calculation. Because of the simple shape of
the Fermi surfaces, one can calculate their volumes from the extremal cros-section, assuming a full
cylindrical symmetry for the Se pocket (which yields $V\approx S_{xz}\sqrt{\pi S_{xy}})$, and using a
trapezoidal integration for the open pockets ($V\approx \lbrack S_{1}+S_{2}]\pi /c).$ Applying this procedure
to Corcoran's numbers we obtain a violation of the Luttinger theorem of the order of more than 1\%, while our
method \cite{CAH} gives the correct total number of electrons with an error of less than 0.01\%.

While ARPES data on the Nb Fermi surfaces agrees well with our calculations, the de Haas-van Alphen area for
the Se pocket (the only one measured so far) differs by nearly 4000 T for the largest orbit and 500 T for the
smallest one. However, due to the large effective mass for this pocket, a relatively minor charge transfer of
only 0.02 electrons from this band to the Nb bands is needed to bring the cross-sections into agreement.  
The geometry of the large Nb cylinders would not be changed in any important way with such a transfer.
Therefore, we can safely use the calculated band structure to evaluate the nesting effects and susceptibility
in NbSe$_{2}$.

\begin{table}[tbp]
\begin{tabular}{|l|ccccc|}
\hline
FS sheet& DOS &
$\omega_{px}$ 
&$\omega_{pz}$ 
&$v_{Fx}$ 
&$v_{Fz}$  \\
& (states/Ry) & (eV) & (eV) & (10$^8$ cm$^{-1}$) & (10$^8$ cm$^{-1}$) \\
 \hline
Band 1 $\Gamma-A$  & 3.8 & 0.40& 2.16& 0.10& 0.53 \\
Band 2 $\Gamma-A$  & 9.9 & 1.63& 0.78& 0.23& 0.11 \\
Band 2 $K-L$       & 19.7& 1.65& 0.86& 0.18& 0.09 \\
Band 3 $\Gamma-A$  &11.4& 1.60& 0.13& 0.22& 0.02\\
Band 3 $K-L$       &29.5& 1.85& 0.26& 0.16& 0.02\\
total              &74.2& 3.37& 2.46&0.19&0.14\\
 \hline
\end{tabular}
\caption{The density of states, plasma frequencies and Fermi velocities for the 5 sheets of
the Fermi surface.}
\label{tr}
\end{table}

The small pancake FS, although it carries little DOS and is probably not important for superconductivity or CDW formation,
plays an important role in transport.  We have calculated the Fermi velocities and plasma frequencies for both in-plane and
perpendicular directions.  A decomposition by band, as shown in Table \ref{tr}), reveals that while the pancake (band 1)
contributes barely 5\% of the total DOS, it is carrying more than three quarters of the total current in the $c$ direction,
and is responsible for the relatively 3D character of the resistivity in this compound. If, as discussed above, we adjust
the relative position of the Se pancake and the other bands, $\omega_{pz}^2$ for this band would drop by less than 50\%, so
that it would still contribute about as much as all other bands together to the transport across the planes. This, has
important ramifications for superconductivity: any experiment related to tunneling (Josephson contacts, Andreev reflection
$etc...$) in the direction perpendicular to the plane will probe mostly the Se pancake band which carries a very small part
of the Cooper pairs and may actually have a reduced superconducting gap (reference to ARPES). Furthermore, if there is any
difference between the superconducting properties of different Nb-derived sheets, this may also manifest itself in a
nontrivial way in such tunneling, since the third band has nearly zero $k_z$ dispersion and and its contribution to the
$c$-transport is nearly negligible (See Table \ref{tr}).

We now discuss the scenarios previously suggested as candidates for nesting at $ \mathbf{Q}$$_{CDW}$ and
compare them quantitatively with our theoretical Fermi surfaces. The six symmetry-equivalent saddle points
along the $\Gamma -K$ directions (not shown) are separated by $\mathbf{Q}_{SP1}$ =
($\frac{1}{6},\frac{1}{6},0$) or by $\mathbf{Q}_{SP2}$ = ($\frac{1}{2},0,0$), neither of which is near
$\mathbf{Q}_{CDW}$. Moreover, the calculated saddle point is approximately 150 meV beneath the Fermi energy,
and although ARPES measurements \cite{TSTF+99} show that band renormalization reduces the depth to only 50
meV, this is still too great an energy compared to the $k_{B}T_{CDW}$=3meV required for the saddle point
theory of Rice and Scott \cite{TMR75}, as has been noted by others \cite{KROS+01}. There have been
suggestions that self-nesting between the parallel flat edges of the central hexagonal Fermi surfaces around
$\Gamma -A$ could be responsible for the CDW instability \cite{TSTF+99}. Our calculations show that only the
smaller (second band) surface is oriented correctly to produce self-nesting along $\Gamma $$-M$; the larger
surface would self-nest along $\Gamma $$-K$. The spanning vector between the faces of the inner surface is
($0.41,0,0$) which is too large in comparison to $\mathbf{Q}$$_{CDW}$ (that of the outer surface is, of
course, even larger). A similar conclusion was reached in Refs. \onlinecite{KROS+01} and
\onlinecite{TKTY+02}. Furthermore, due to substantial $k_{z}$ dispersion in the second pocket, good nesting
does not occur along the full length of the cylinder, substantially reducing $L_{k}$ in Eq. \ref{nesting} and
thereby diminishing the contribution to $\chi $. Rossnagel \textit{et al} \cite{KROS+01}, noting this
dispersion along $k_{z}$, suggested that the broad peak calculated \cite{NJD78,KMEA83} in $\chi
_{0}(\mathbf{Q}_{CDW},0)$ is due to imperfect (weak) nesting between these warped cylindrical surfaces
themselves (self-nesting) and between neighboring cylinders. Our calculation of $\chi ({\mathbf{q},0})$ shows
that such a scenario is not manifested for NbSe$_{2}$ (see Section III). It has been recently postulated
\cite{TVAVF+04} that the $K-L$ centered cylinders, which are smaller than those around $\Gamma -A$, might
nest at approximately the right wave vector. The cross-sections of these cylinders, however, are not circles
but are rounded triangles with flat edges oriented at 120$^{\circ }$ with respect to one another, making such
nesting highly unlikely. The only apparent strong nesting possibility, according to our calculations occurs
between the flat triangular edges of the $K-L$ cylinders and the parallel flat edges of the central hexagonal
pocket, $\mathbf{Q}$ = ($\frac{1}{3},\frac{1}{3},0$). In the next section, we will show that it is indeed
this nesting that produces the highest peaks in the imaginary part of the susceptiblity. We reiterate here
that FS nesting causes peaks in the imaginary part of $\chi _{0}$, while the real part of $\chi _{0}$, which
is the essential quantity for CDW formation, draws from a much larger energy range and may have features
entirely unrelated to FS topology.

\section{Non-Interacting Susceptibility}

\begin{figure}[tbp]
\includegraphics[width=0.9\linewidth]{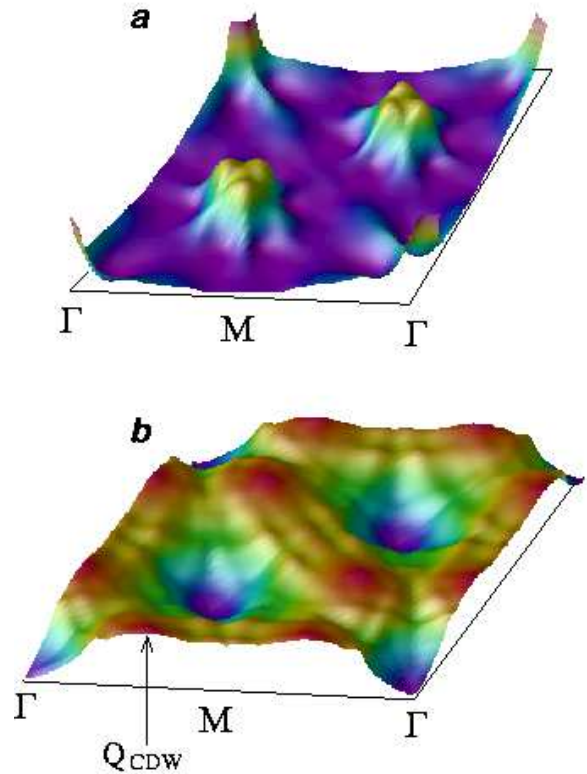}
\caption{The non-interacting susceptibility of NbSe$_2$. a) The imaginary
part exhibits FS nesting driven peaks at $\mathbf{Q}$ = ($\frac{1}{3},\frac{1
}{3},0$). The plane at the bottom is a guide for the eye and corresponds to
the lowest value of $\Im\protect\chi_0$. b) The real part has very weak
peaks at $\mathbf{Q}_{CDW}$ that come from energy intervals away from E$_F$.
(see text). The plane corresponds to the lowest value of $\Re\protect\chi_0$
, which is also the density of states at the Fermi level.}
\label{susc}
\end{figure}

The calculated real and imaginary parts of the non-interacting susceptibility are shown in Fig.
\ref{susc}. The matrix elements of Eq. \ref {Re} have been set to $\delta_{ll'}$ so that no interband
contributions are taken into account.  The effects of these terms turn out to be quite weak as we will
show later in the text.  The clear peaks in $\Im\chi_0$ indicate nesting at $ \mathbf{q}$ =
($\frac{1}{3},\frac{1}{3},0$) and by symmetry, at $\mathbf{q}$ = ($\frac{2}{3},\frac{2}{3},0$). These
correspond to translating the central pocket from $\Gamma$ to $K$. This results in partial nesting
between the hexagonal and triangular pockets, as well as between two triangular pockets with different
orientations. Since the pockets can slide around inside one another, the peak is broad and, in fact, is
not truly a single peak, but a composite peak composed of three separate peaks \textit{near}
$\mathbf{q}=K$, and corresponding to separate nestings of the three edges of each triangle. Assuming
that the triangular surfaces are, in actuality, more rounded as indicated by ARPES, we expect these
three peaks to further ``smear'' into one central peak. There is no indication of any other nesting
and, most significantly, there is no peak whatsoever at $\mathbf{q} = \mathbf{Q}_{CDW}$. This situation
differs from that of isostructural NbTe$_2$ where a strong peak in $\Im\chi({\mathbf{q},0})$ does exist
\cite{CBJC+05} at $\mathbf{Q} _{CDW}$, though we note that such a peak cannot be taken as evidence that
nesting contributes to CDW formation without a calculation of $\Re\chi({\mathbf{q},0})$. Unless the
true fermiology of NbSe$_2$ differs signficantly from our calculations (and comparison with experiment
strongly suggests that it does not), there is \textit{no evidence} of a FS nesting contribution to CDW
formation at all.

\begin{figure}[tbp] \includegraphics[width=0.99\linewidth]{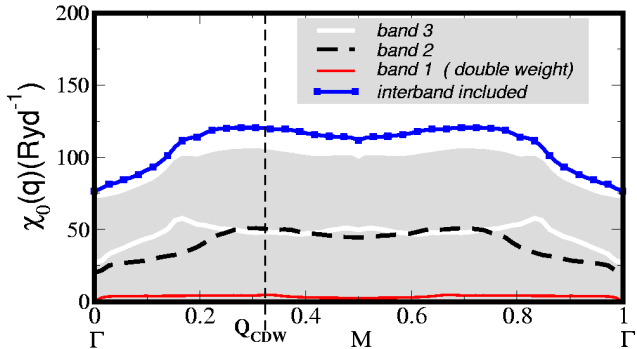} \caption{$\Re\protect\chi_0(\mathbf{q})$
along the $\Gamma-M$ line of the Brillouin zone. The total is shown in shaded grey with the contributions
from each individual band delineated by different lines.  The height of band 1 has been multiplied by two to
make it visible above the x-axis. The topmost line includes non-diagonal (interband) contributions.}
\label{GM} \end{figure}

The peaks in the imaginary part of the susceptibility are always present in
the real part because $E_{F}$ is part of the sampled energy range, but these
can be either emphasized or overshadowed by contributions from other energies.
In the case of NbSe$_{2}$, the peaks at $K$ are no longer even visible and broad
maxima appear instead along the $\Gamma -M$ line. These do occur
approximately at $\mathbf{Q}$$_{CDW}$, but are unlikely to be strong enough
to produce a CDW instability. A two-dimensional plot of $\Re \chi _{0}$
along $\Gamma -M$ is shown in Fig. \ref{GM}, with the position of the
observed CDW wave vector indicated. The highest point along this curve is
33.5 Ryd$^{-1}$ above the lowest point, $\Re \chi _{0}(0)$=N(E$_{F}$)=70.3
Ryd$^{-1}$, and occurs at $\mathbf{q}$=(0.31,0,0) which is, in fact, rather
near $\mathbf{Q}_{CDW}$. Fig. \ref{GM} includes a band-by-band decomposition
of the contributions to $\Re \chi _{0}(\mathbf{q})$ along $\Gamma -M$
showing almost no contribution at all from the Se-derived band and a rather
peaked contribution from the antibonding Nb band, but at a wave vector quite
different from $\mathbf{Q}_{CDW}$. Finally, we have included a plot of $\Re
\chi _{0}(\mathbf{q})$ with both intra- and interband matrix elements set to
unity in Eq. \ref{Re}. This exaggerates the contributions of interband
transitions, but gives a rough idea of their effect on the susceptibility.
We find an overall shift upward of the entire spectrum but no particular
sharpening of the peaks. Along $\Gamma -M$, our susceptibility somewhat
resembles that of Ref. \onlinecite{NJD+78}, but the suggestion that the broad peak
is due to self-nesting of the central pocket is clearly unjustified as no
such peak appears in $\Im \chi _{0}$.

To isolate the true source of the (weak) maxima, we calculate a function,

\begin{equation}
F(x,y) = \int \delta({\epsilon_{\mathbf{k}}}-x)\delta({\epsilon_{\mathbf{k}+\mathbf{q}
}}-y) d\mathbf{k},
\end{equation}
from which the susceptibility can be recovered through 
\begin{equation}
\chi(\mathbf{q}) = \int_{-\infty}^{\epsilon_F}dx \int_{\epsilon_F}^{\infty}dy 
\frac{F(x,y)}{x-y}  \label{f}
\end{equation}

By adjusting the limits of the integrals in Eq. \ref{f}, the energy interval
that gives rise to any given feature in $\chi_0$ can be isolated. For a very
small interval near the Fermi energy, all the features of $\Im \chi_0$ are
reproduced, including the dominant feature at $K$. However, for energy
intervals away from the Fermi energy, there is little or no contribution to $
\chi_0(K)$, with the result that the real part of $\chi_0$ has a minimum
there when all energies are taken into account. Much of the contribution to $
\Re\chi_0$ comes from eigenvalues in the energy range $|\epsilon_{\mathbf{k+q
}}-\epsilon_{\mathbf{k}}| >$ 0.02 Ryd. This reinforces in a rather dramatic
way our earlier assertion that one should not expect a CDW instability to
stem from FS nesting.

\section{Conclusions}

We have calculated the real and imaginary parts of the non-interacting susceptibility as a function of wave vector
throughout the Brillouin zone. The real part of $\chi _{0}$ exhibits very weak peaks near $\mathbf{Q}_{CDW}$ that are
unlikely to be strong enough alone to cause a CDW instability. The main contributions to these peaks come from an energy
range \textit{not} near E$_{F}$, indicating that FS nesting is irrelevant. A direct calculation of $\Im \chi
_{0}(\mathbf{q})$ confirms that there is absolutely no FS nesting at $\mathbf{Q}_{CDW}$, although peaks due to nesting
along $\mathbf{q}$ = ($\frac{1}{3},\frac{1}{3},0$) are evident. It can therefore be definitively stated that Fermi
surface nesting contributes nothing to CDW instability in NbSe$_{2}$. This is indicative of the more general, but often
overlooked fact that Fermi surface nesting has an immediate effect only on the imaginary part of susceptibility,
relevant, for instance, for the spin excitation spectrum, but not to CDW formation. The latter is defined by the
structure of the real part of susceptibility, which is not directly affected by the Fermi surface nesting. More
precisely, the effect of nesting on $\Re \chi _{0}(\mathbf{q},\omega )$ occurs in a very small part of the total relevant
energy range (which is of the order of electronvolts rather than Kelvins), and if the electronic structure changes
qualitatively within this range for a given system, which is nearly always the case, then nesting cannot be considered a
primary mechanism of CDW formation.

\acknowledgements
We would like to acknowledge several helpful and enlightening discussions with Girsh Blumberg and Aleksej 
Mialitsin.  Research at NRL is supported by the Office of Naval Research.

\end{document}